\documentclass[aps,prl,twocolumn,groupedaddress,showpacs]{revtex4}
\usepackage{natbib}
\usepackage{graphicx}

\bibliography{12}
\input{tcilatex}

\begin{document}

\title{Resonances in 1D disordered systems:\\
localization of energy and resonant transmission}
\author{K.Yu.Bliokh}
\affiliation{Institute of Radio Astronomy, Kharkov 61002, Ukraine}
\author{Yu.P.Bliokh}
\affiliation{Department of Physics, Technion, Haifa 32000, Israel}
\author{V.Freilikher}
\affiliation{Department of Physics, Bar-Ilan University, Ramat-Gan 52900, Israel}

\begin{abstract}
Localized states in one-dimensional open disordered systems and their
connection to the internal structure of random samples have been studied. It
is shown that the localization of energy and anomalously high transmission
associated with these states are due to the existence inside the sample of a
transparent (for a given resonant frequency) segment with the minimal size
of order of the localization length. A mapping of the stochastic scattering
problem in hand onto a deterministic quantum problem is developed. It is
shown that there is no one-to-one correspondence between the localization
and high transparency: only small part of localized modes provides the
transmission coefficient close to one. The maximal transmission is provided
by the modes that are localized in the center, while the highest energy
concentration takes place in cavities shifted towards the input. An
algorithm is proposed to estimate the position of an effective resonant
cavity and its pumping rate by measuring the resonant transmission
coefficient. The validity of the analytical results have been checked by
extensive numerical simulations and wavelet analysis.
\end{abstract}

\pacs{72.15.Rn, 42.25.Dd, 72.10.Fk}
\maketitle

\address{$^1$Institute of Radio Astronomy, Kharkov 61002, Ukraine,\\
         $^2$Department of Physics, Technion, Haifa 32000, Israel,\\
         $^3$Department of Physics, Bar-Ilan University, Ramat-Gan 52900, Israel}


\section{Introduction}

Localization of waves in random media has been investigated intensively
during last few decades. One-dimensional (1D) strong localization has
received the most study, both analytically and numerically. Localization of
eigen states in closed 1D disordered systems and exponentially small (with
respect to the length) transparency of open systems with 1D disorder have
been studied with mathematical rigor (see, for example, Ref. 1 and
references therein). The most common physical manifestation of these effects
is the fact that a thick stack of transparencies reflects light as a rather
good mirror.\cite{berry} Much less evident (though also well known
theoretically) is that for each (sufficiently long ) disordered 1D sample,
there exists a quasi-discrete set of frequencies that travel through the
sample unreflected, i.e. with the transmission coefficient close to one.$%
^{3-5}$ Values of these frequencies are random, determined exclusively by
the structure of the realization, and the whole set of frequencies
constitutes a fingerprint of a given random sample.

High transparency is always accompanied by a large concentration
(localization) of energy in randomly arranged points inside the system.
Therefore, an open random 1D sample can be considered as a resonator with a
set of modes (resonances) with high quality factors. An important feature of
such systems is that, in distinction to a regular resonator whose modes
occupy all inner space, in a 1D random sample each mode (frequency) is
localized inside its own effective 'cavity' whose position is random and
size is much smaller than the size of the sample. The resonances are sharp
and well pronounced in the sense that their widths are much smaller than the
distances between them, both in the frequency domain and in real space.
Therefore, it is of little wonder that resonances caused by one-dimensional
disorder have recently attracted considerable interest of physicists looking
for the possibility of creating random laser.$^{6-10}$

In this paper, we investigate the localized states (modes, or resonances) in
one-dimensional open random systems. To understand the nature of the
resonances in disordered 1D systems and to describe them quantitatively, the
problem is mapped onto a quantum mechanical problem of tunnelling and
resonant transmission through an effective two-humped regular potential.
With the formulas derived on the basis of this mapping, we find statistical
characteristics of the resonances, namely, their spectral density, spectral
and spatial widths, as well as the amplitudes of the field peaks and
transmission coefficient. Based on these results, one can solve an inverse
problem, namely, to predict (with some probability) for each resonance the
position of the effective cavity and the maximal pumping amplitude, using
the total length of the sample and localization length as the fitting
parameters, and the transmission coefficient as the only (measurable) input
datum. The analytical results deduced from the quantum-mechanical analogy
are in close agreement with the results of numerical experiments.

We study the frequency dependence of the transmission coefficient and the
spatial intensity distribution induced by incident monochromatic waves with
different frequencies. If the frequency of the incident wave coincides with
one of the 'eigen frequencies' of the sample, the energy is genuinely
localized in some randomly located area of the size of the localization
length. Thus, the transparency of this area is abnormally high, whereas the
transparency of the adjacent segments is exponentially small. The
transparent segment surrounded by non-transparent parts of the sample
constitutes an effective resonant cavity with a high Q-factor. The
transmission coefficient at a resonance is independent of the total length, $%
L$, of the sample and is determined only by the location of the cavity, that
is, totally different from the typical transmission that decays
exponentially with increasing $L$ . The maximal transmission is provided by
the modes that are located in the center, while the highest energy
concentration takes place in cavities shifted towards the input. The number
of the resonantly transparent frequencies is shown to be $L/l_{loc}\gg 1$
times smaller than the total number of resonances ($l_{loc}$ is the
localization length). Wavelet analysis of random realizations supports the
validity of the mapping and provides physical insight into the origin of the
effective cavities in disordered samples.

\section{Formulation of the problem}

We consider a plane monochromatic wave with wavelength $\lambda $ and unite
amplitude incident from the left ($x<0$) on a 1D disordered sample with
randomly fluctuating refractive index. The transmission coefficient, $%
T(\lambda )$ and the random field amplitude, $A=|E(x)|$, have been studied
both analytically and numerically. Typical frequency dependence of the
transmission coefficient of a 1D randomly layered sample is shown in Fig.1.

One can see that along with a continuum of wavelengths for which the
transmission coefficient is exponentially small ($\sim \exp (-2L/l_{loc})$),
there is a discrete set of points at $\lambda$-axis ($\lambda _{b},\lambda
_{c}$ in Fig.1) where $T(\lambda )$ has well pronounced narrow maxima. The
amplitude distributions induced by the corresponding waves inside the sample
are shown in Fig.2. While at typical frequencies (realizations) the
amplitude of the field decrees exponentially from the input (Fig. 2, $%
\lambda _{a}$), the resonances, (Fig. 2, $\lambda _{b,}$ $\lambda _{c}$),
exhibit essentially non-monotonous spatial distribution of the amplitude.

\begin{figure}[htb]
\centering \includegraphics[width=80mm,height=60mm,angle=0,]{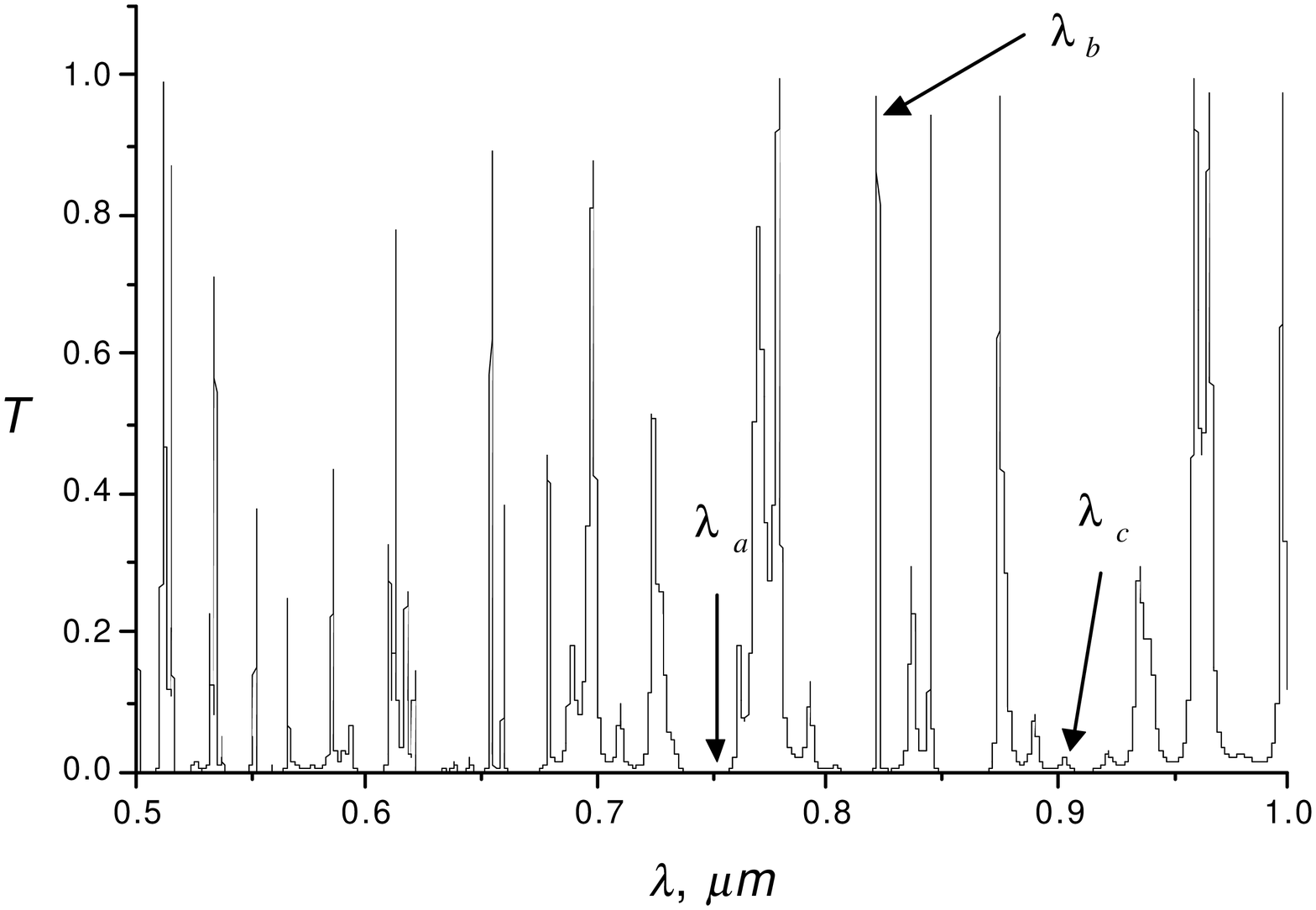}
\caption{Transmission coefficient as a function of the wavelength.}
\end{figure}

\begin{figure}[htb]
\centering \includegraphics[width=80mm,height=60mm,angle=0,]{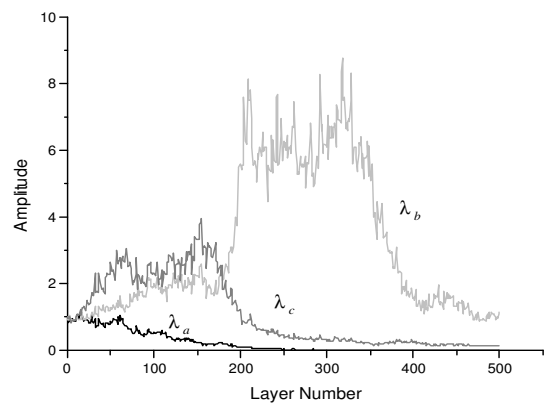}
\caption{Amplitude of the field inside the sample as a function of the
coordinate for three wavelengths marked in Fig. 1.}
\end{figure}

Important to note that the localization of energy takes place for all
resonances, not only for those with the transmission coefficient close to
one (Fig. 2, $\lambda _{b,}$ $\lambda _{c}$). The amplitude of a maximum
depends on its location in space, which in its turn, is uniquely determined
by the internal structure of the realization. As it is shown below, this
fact provides a means for evaluating the resonant amplitude and the
coordinate of the point where the resonant mode is localized, if the total
transmission coefficient is known.

\begin{figure}[tbh]
\centering\includegraphics[width=80mm,height=100mm,angle=0,]{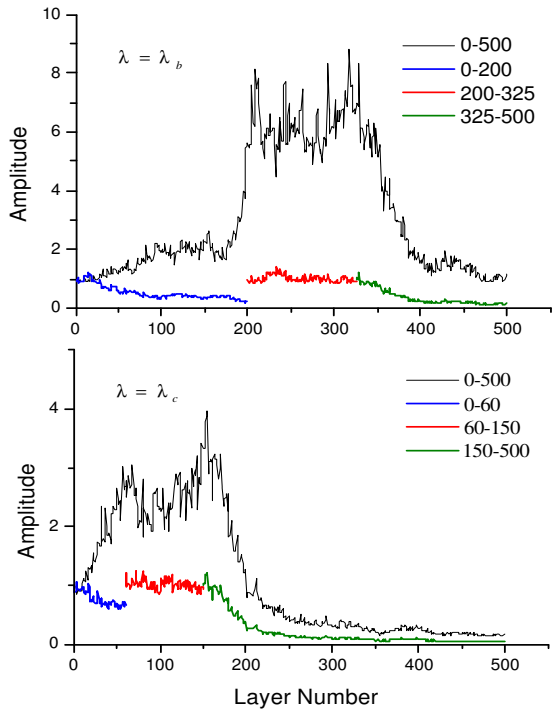}
\caption{Amplitude of the field as a function of the coordinate inside the
whole sample (black curves), in the left (blue curves), middle (red curves),
and right (green curves) parts taken as a separate sample each. $\protect%
\lambda =\protect\lambda _{b}$ and $\protect\lambda =$ $\protect\lambda _{c}$%
.}
\end{figure}

Fig. 3 demonstrates the connection between the amplitude distribution and
the transparency of different parts of the sample. The upper picture shows
the intensities of the field generated by a resonantly transmitted wave with 
$T\simeq 1$ ($\lambda _{b}$ in Fig. 1) inside the whole sample (black curve)%
\textit{, }and in their left (blue curve), middle (red curve), and right
(green curve) parts when they are taken as a separate sample each. The same
dependencies for resonant wave with $T<1$ ($\lambda _{c}$ in Fig. 1) are
depicted in the lower picture. It is seen that the middle sections where the
energy is concentrated are almost transparent for the wave, while the side
parts are practically opaque for the resonant frequencies. Wavelet analysis
of the power spectra of the transparent and non-transparent parts presented
in Sec. 6 provides physical insight into their nature.

Structures of this type have been studied in the quantum mechanical problem
of the tunnelling and resonant transmission of particles trough a regular
(non random) potential profile consisted of a well bounded at both sides by
two potential barriers (Fig. 4) \cite{bohm}. Although the physics of the
propagation in each system is totally different (interference of the
multiply scattered random fields in a randomly layered medium, and
tunnelling through a regular two-humped potential), the similarity of this
two problems turns out to be rather close. Indeed, in both cases the
transmission coefficients are exponentially small for most of frequencies
(energies), and have well-pronounced resonant maxima (sometimes of order of
unity) at discrete points corresponding to the eigen levels of each system.
The energy at resonant frequencies is localized in a transparent part, and
the total transmission depends drastically on the position of this part.
More than that, even qualitatively the intensity distributions presented in
Fig. 3 and the corresponding values of the transmission coefficients compare
favorably with those calculated for a potential profile depicted in Fig. 4
if the parameters of the effective profile have been chosen properly. 
\begin{figure}[htb]
\centering \includegraphics[width=80mm,height=60mm,angle=0,]{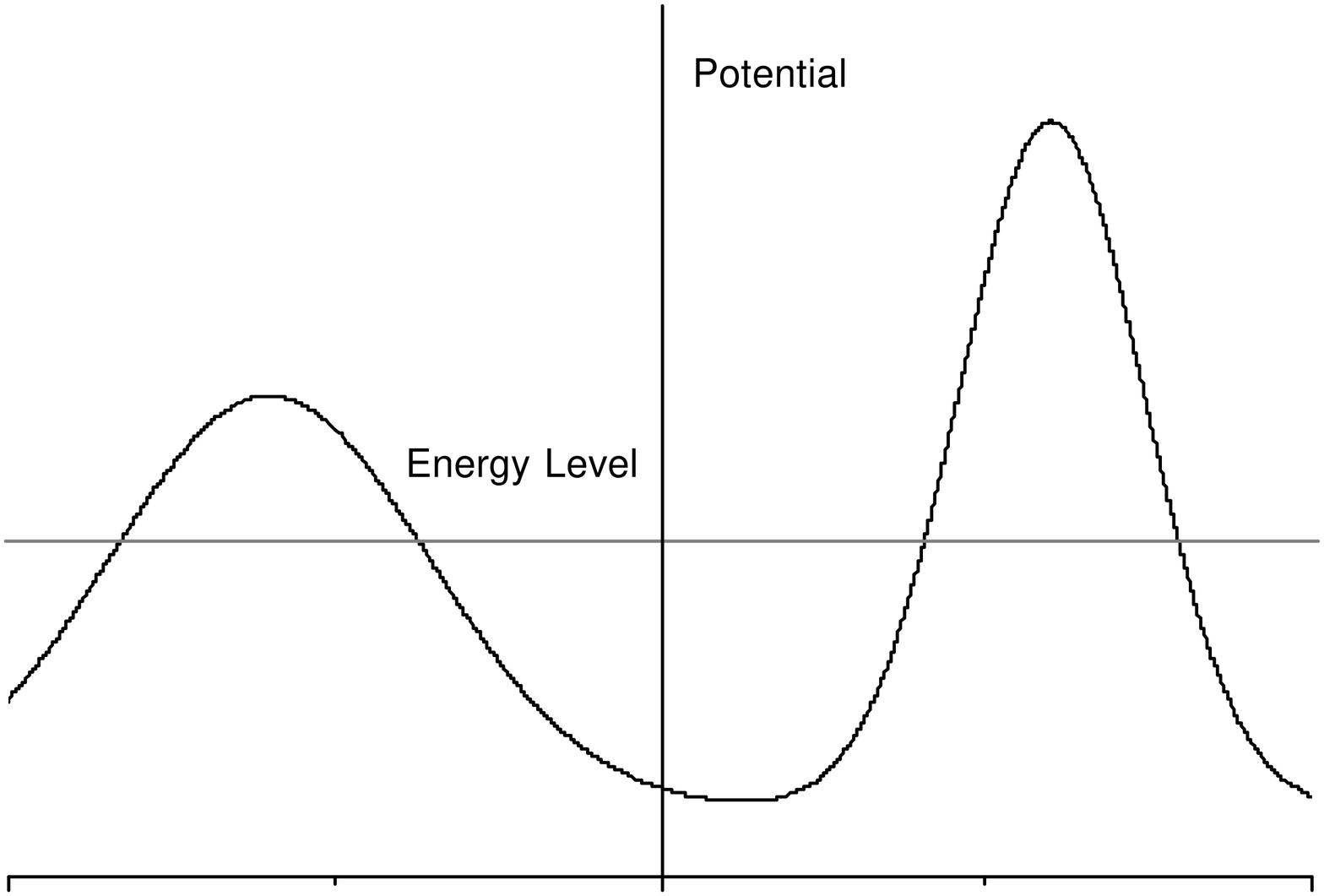}
\caption{Two-humped potential profile.}
\end{figure}

\section{Quantum mechanical formulas}

To map the stochastic classical wave scattering problem in hand onto a
deterministic quantum mechanical one, that is to say, to construct an
appropriate effective regular potential that provides the same transmission
and the interior intensity distribution, let us consider an auxiliary
problem of the transmission of a quantum particle through a potential
profile (resonator) consisted of two barriers separated by a potential well
(Fig. 4). The transmission coefficient through the potential can be
calculated in WKB approximation \cite{bohm}, which yields 
\begin{widetext}
\begin{equation}
T=\left[ \left( \frac{\Theta _{1}^{2}+\Theta _{2}^{2}}{2\Theta _{1}\Theta
_{2}}\right) \cos ^{2}\frac{1}{2}(\pi -J)+\frac{1}{4}\left( 4\Theta
_{1}\Theta _{2}+\frac{1}{4\Theta _{1}\Theta _{2}}\right) ^{2}\sin ^{2}\frac{1%
}{2}(\pi -J)\right] ^{-1}~.
\end{equation}
\end{widetext}
Here $\Theta _{i}=T_{i}^{-1/2}\gg 1$ with $T_{i}$ being the tunnelling
transmission coefficients of the barriers, $J/2$ is the phase acquired in
the well, 
\begin{equation}
J=2\int_{l_{0}}k(x)dx\sim 2kl_{0}~,
\end{equation}
$k(x)$ is the momentum ($\hbar =1$), and $l_{0}$ is the length of the well.
At $\Theta _{1}=\Theta _{2}\equiv \Theta $ (symmetrical potential) equation
(1) transforms into 
\begin{equation}
T=\left[ 1+\frac{1}{4}\left( 4\Theta ^{2}-\frac{1}{4\Theta ^{2}}\right)
^{2}\sin ^{2}\frac{1}{2}(\pi -J)\right] ^{-1}~.
\end{equation}
Since $\Theta _{i}^{2}\gg 1$, it follows from Eqs. (1), (3) that typically $%
T\ll 1$. Resonance transmission ($T\simeq 1$) takes place only at the
discrete points that correspond to the eigen levels of the well: 
\begin{equation}
J=\pi \left( n+\frac{1}{2}\right) ~,~~n=0,1,2,...~.
\end{equation}%
Thus, the characteristic spectral interval between the resonances is
determined by the requirement $\Delta J=\pi$, or 
\begin{equation}
\Delta k_{res}\sim \frac{1}{l_{0}}~.
\end{equation}

To estimate the half-width of the resonances, $\delta k$, we note that in
the vicinity of $k_{res}$ the transmission coefficient Eq. (3) takes the
form 
\begin{equation}
T\simeq\frac{1}{1+4l_{0}^{2}\left(k-k_{res}\right)^2\Theta^4}
\end{equation}
(with $T_{res}=1$), from which it follows that 
\begin{equation}
\delta k\simeq\frac{1}{2l_0\Theta^2}~.
\end{equation}

In general (asymmetrical) case Eq. (1) the transmission coefficient at a
resonance is equal 
\begin{equation}
T_{res}=\left( \frac{2\Theta _{1}\Theta _{2}}{\Theta _{1}^{2}+\Theta _{2}^{2}%
}\right) ^{2}~.
\end{equation}
It is easy to see that asymmetry of the potential profile reduces
drastically the resonance transmission: $T_{res}\ll 1$ if $\Theta_{1}^{2}\ll 
$ $\Theta _{2}^{2}$ or $\Theta _{1}^{2}\gg \Theta _{2}^{2}$. In the same
approximation the amplitude of the wave function in the well is given by 
\begin{equation}
|A|\simeq \frac{2\sqrt{2}\Theta _{1}\Theta _{2}^{2}}{\Theta
_{1}^{2}+\Theta_{2}^{2}}.
\end{equation}

\section{Application to the resonances in disordered systems}

To use the above-derived formulas for the qualitative description of the
wave transmission through a random sample we have to express the parameters
of the effective potential through the statistical parameters of the
disordered system in hand. For an ensemble of 1D random realizations those
parameters are the total length $L$ and the localization length $l_{loc}$
that we define as $l_{loc}=2L/<\ln T^{-1}>^{-1}$. Note that in the localized
regime $L\gg l_{loc}\gg \lambda $. This inequality justifies the validity of
WKB approximation. To determine properly the length of the (effective) well
in disordered systems we note that the appearance of a transparent segment
(effective well) inside a random sample is the result of a very specific
(and therefore low-probable) combination of phase relations. Obviously, the
longer such segment is, the less is the probability of its occurrence. On
the other hand, the typical scale in the localized regime is $l_{loc}$.
Hence, the minimal and, therefore, the most probable size of the effective
well is $l_{0}\sim l_{loc}$ that we assume to be the same in all
realizations. Under this assumption, different values of the resonant
transmission coefficients and different intensity distributions at different
resonant realizations can be reproduced by variations of the location of the
well in the corresponding quantum formulas presented in Sec. 3 with 
\begin{equation}
l_{0}=l_{loc}~.
\end{equation}

If the center of the transparent segment of a resonant realization is
shifted on the distance $d$ from the center of the sample, the lengths of
the non-transparent parts of resonant realizations are 
\begin{equation}
l_{1,2} =\frac{L-l_{loc}}{2}\pm d~,
\end{equation}
with transmission coefficients $T_i=\exp\left(-2l_i/l_{loc}\right)$.
Therefore functions $\Theta_i$ in Eqs. (1)-(9) should be taken in the form 
\begin{equation}
\Theta _{1,2}=\exp\left( \frac{l_{1,2}}{l_{loc}}\right) =\exp \left( \frac{%
L-l_{loc}}{2l_{loc}}\pm \frac{d}{l_{loc}}\right) ~.
\end{equation}

The above-introduced mapping, Eqs. (10)-(12), allows the use of Eqs.(1)-(9)
for evaluating the corresponding quantities related to random 1D systems. It
becomes apparent, for example, that not all eigen modes of along random
sample provide high transmission ($T\simeq 1$) through the system. Indeed,
while the distance (in $k$-space) between eigen modes is 
\begin{equation}
\Delta k\sim \frac{1}{L}~,
\end{equation}%
the typical interval between the resonantly transparent frequencies obtained
from Eq. (5) by substituting Eq. (10) is of order 
\begin{equation}
\Delta k_{res}\sim \frac{1}{l_{loc}}~.
\end{equation}%
It is easy to see that $\Delta k/\Delta k_{res}\sim l_{loc}/L\ll 1$, which
means that the resonances with $T\sim 1$ occur much less frequently than all
other resonances, $T<1$. It is physically clear, and follows from the fact
that a resonance with high transmission occurs when the transparent segment
is located in the middle of the sample (with accuracy of $l_{loc}$), while
the resonances with smaller transmission are observed in any non-symmetrical
potentials with arbitrary situated transparent segment. Assuming that the
transparent segment can be found at any point of the sample with equal
probability, we infer that resonances with $T\sim 1$ are encountered $%
L/l_{loc}$ times more rarely than all other ones.

It follows from Eq. (7) that typical width of resonances is 
\begin{equation}
\delta k\sim \frac{1}{l_{loc}}\exp \left( -\frac{L}{l_{loc}}\right) ~,
\end{equation}
that is to say, it decreases exponentially with the length of the sample.

In the same manner, the resonant transmission coefficient at resonant
frequencies and peak amplitudes can be estimated from Eqs. (8) and (9),which
after substituting Eq. (12) yield: 
\begin{equation}
T_{res}(d)=\frac{4}{\exp (2d/l_{loc})+\exp (-2d/l_{loc})}~,
\end{equation}
\begin{equation}
|A(d)|^{2}=\frac{8\exp (L/l_{loc}-1-2d/l_{loc})}{\left[ \exp
(2d/l_{loc})+\exp (-2d/l_{loc})\right] ^{2}}.
\end{equation}
Eq. (16) shows that the resonant transmission coefficients do not depend on
the total length of the sample and are governed only by the positions of the
areas of localization, in contrast to the typical transmission which decays
exponentially with $L$ increasing.

If $T_{res}\simeq 1$, then $d\simeq 0$, and Eq. (17) gives 
\begin{equation}
|A|\simeq \sqrt{2}\exp \left( L/2l_{loc}\right) \gg 1~.
\end{equation}
When the transparent segment of a random realization is shifted from the
center, it follows from Eq. (17) that 
\begin{widetext}
\begin{equation}
|A|\simeq \left[
\begin{array}{cl}
2\sqrt{2}\exp (l_{1}/l_{loc})\gg 1, & \text{if}~ \exp(l_{1}/l_{loc})\ll \exp
(l_{2}/l_{loc}) \\
2\sqrt{2}\exp \left( (2l_{2}-l_{1})/l_{loc}\right) \gg 1, & \text{if}~
\exp(2l_{2}/l_{loc})\gg \exp (l_{1}/l_{loc})l_{1}\gg \exp (l_{2}/l_{loc}) \\
2\sqrt{2}\exp \left( (2l_{2}-l_{1})/l_{loc}\right) \ll 1, & \text{if}~
\exp(l_{1}/l_{loc})\gg \exp (2l_{2}/l_{loc})%
\end{array}
\right.
\end{equation}
\end{widetext} with $l_{1,2}$ defined in (11).

It can be shown from Eq. (17) that the maximum of the amplitude is reached
when the transparent part is shifted from the center of the sample towards
the input on the distance 
\begin{equation}
d=-\frac{1}{4}\ln \frac{1}{3}~l_{loc}\simeq -0.27l_{loc}~.
\end{equation}%
This shift is independent of the length of the sample.

\section{Numerical simulations}

To test the validity of the above-introduced analytical results, Eqs.
(13)-(20), numerical calculations of the spectrum of resonances and of the
spatial intensity distributions at resonant frequencies have been performed
for more than $10^{4}$ resonances. In the calculations we consider samples
with up to $N=1000$ of layers, and assume that the refractive indexes and
sizes of the layers are independent random variables uniformly distributed
in the ranges $n=1\pm 0.25$ and $d=0.15\pm 0.05~\mu m$ respectively. The
wavelength varies in the interval $0.5\mu m\leq \lambda \leq 1.5\mu m$.

In accordance with Eq. (14) the average spacing between resonantly
transparent modes with $T\simeq 1$ depicted in Fig. 5 (curve b) is
determined by the localization length and practically independent on the
size $L$ of the system. For comparison, the numerically calculated $L$%
-dependence of the eigen mode spacing is also shown in Fig. 5 (curve a). It
fits well Eq. (13). Numerically calculated average half-width of the
resonances (Fig. 6) is consistent with the analytical expression Eq. (15).
The localization length had been computed through the transmission
coefficient as $l_{loc}=-2L<\ln T>.$ Note that, being a self-averaging
quantity \cite{lif}, $l_{loc}$ slightly fluctuates from sample to sample,
and can be estimated from the transmission coefficient at a typical
realization.

\begin{figure}[htb]
\centering \includegraphics[width=80mm,height=90mm,angle=0,]{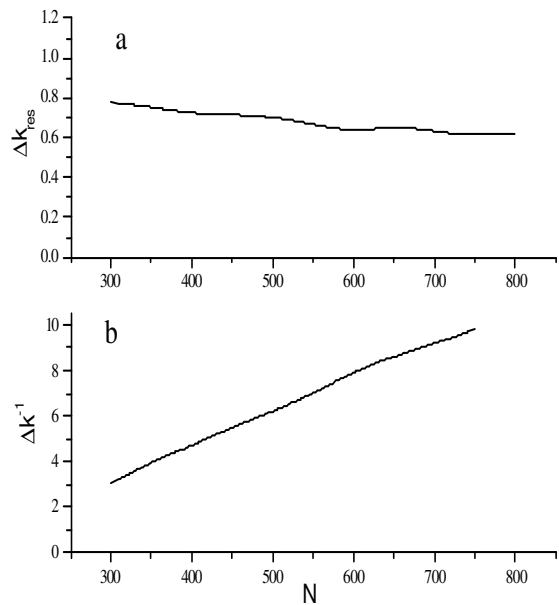}
\caption{Inverse spacing between eigen modes, (a), and spacing between
resonantly transparent eigen modes, $T\simeq 1,$ (b), as functions of the
length of the sample.}
\end{figure}

\begin{figure}[htb]
\centering \includegraphics[width=80mm,height=60mm,angle=0,]{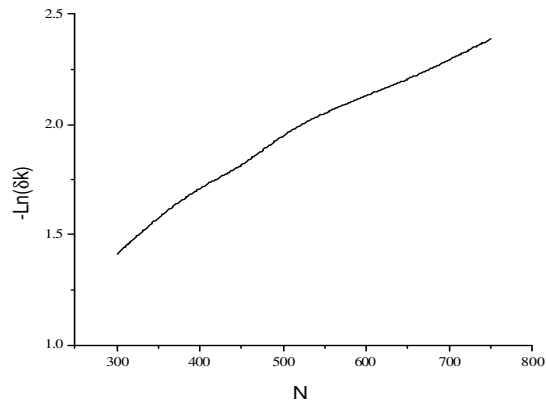}
\caption{Logarithm of the half-width of the resonances as a function of the
length of the sample.}
\end{figure}

Depicted in Fig. 7 is the probability that a mode localized at a given point
provides transmission coefficient $T$ (dimensionless coordinate $x/L$ is
used). As is seen from the picture, the probability of high transitivity ($%
T\simeq 1$) has maximum for modes that are localized in the center of the
random sample. As the location of an effective cavity shifts from the
center, the corresponding transmission coefficient decreases, in agreement
with Eq. (16). To make the comparison with the analytical results more
convenient, the same (as in Fig. 7) probability density distribution is
presented in Fig. 8 as a two-dimensional picture where different colors
correspond to different probabilities. Black line displays the transmission
coefficient calculated by Eq. (16) as a function of the coordinate of the
corresponding point of localization. Note that shifted to the exit peaks of
the field are not shown in Fig. 8. They are much weaker, because they happen
on the background of an exponentially small signal.

\begin{figure}[htb]
\centering \includegraphics[width=80mm,height=60mm,angle=0,]{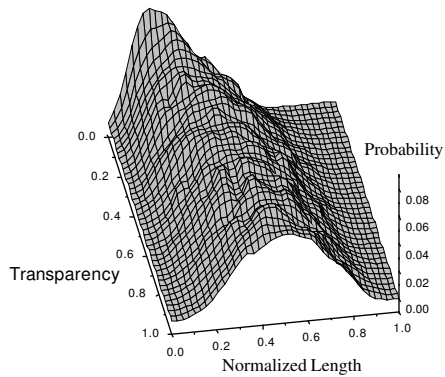}
\caption{Probability for a mode localized at a given point to provide the
value of the transmission coefficient $T$ , as a function of the
dimensionless coordinate $x/L.$}
\end{figure}

\begin{figure}[htb]
\centering \includegraphics[width=80mm,height=60mm,angle=0,]{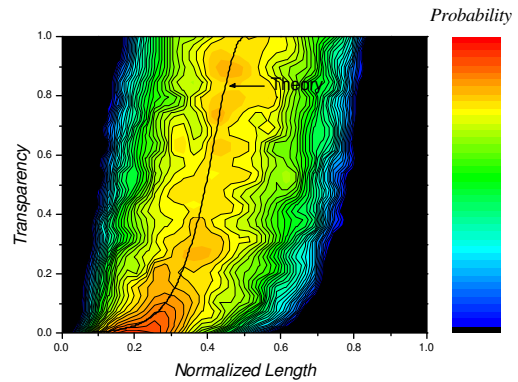}
\caption{ The same probability as in Fig. 7 presented by colors. Black line
displays the transmission coefficient as a function of the coordinate of the
point of localization calculated by Eq. (16).}
\end{figure}

Numerically calculated average normalized field amplitude of localized modes
(pumping rate of effective resonant cavities) as a function of the
coordinate of the cavity exponentially increases with the length of the
sample in a good agreement with Eq. (17). Effective cavities which provide
the highest pumping rate are shifted from the center towards the input in
accordance with Eq. (20). Interestingly enough, the seemingly rough analogy
based on only one fitting parameter (localization length) performs
surprisingly well. Indeed, not only the shift is independent on the total
length and proportional to $l_{loc}$ as predicted by Eq. (20), but also the
coefficient in Eq. (20) coincides with that obtained in numerical
simulations with the accuracy 10\%. We have also verified that the maximal
peak amplitudes exponentially increase with the length of a sample and by
the order of magnitude coincide with the values given by Eqs. (18).

\section{Wavelet analysis of random samples}

Here we discuss briefly why the effective resonant cavities exist in 1D
disordered samples, in other words, why different segments of a random
sample are transparent or non-transparent for the wave with given wave
number $k={2\pi }/{\lambda }$. It is well known that in the case of a weak
scattering the so called resonance reflection takes place, that is, the
reflection coefficient (and, therefore, the transparency) of some segment of
a random medium is determined by the amplitude of the resonance harmonic
with $q_{res}=2k$ in the spatial Fourier transformation of its refractive
index \cite{ryt}. In other words, any long enough randomly layered segment, $%
x_{1}<x<x_{2},$ resonantly reflects as a periodically modulated medium whose
modulation wave number $q$ is determined by the resonance condition $%
q=q_{res}=2k,$ and the amplitude, $\widetilde{n}^{2}(q_{res})$, is 
\begin{equation}
\widetilde{n}^{2}(q_{res})=\int_{x_{1}}^{x_{2}}n^{2}(x)e^{-i2kx}dx~.
\end{equation}
This expression is known as window Fourier transform with the rectangular
window function, and characterizes the local spectrum of the investigated
function. The rectangular window function has some disadvantages that
disappear when instead of the window Fourier transform, a wavelet transform
is applied for the determination of the local spatial spectrum of the
sample. To investigate the local spectra of different parts of the sample we
used the so-called complex-valued Morlet wavelet \cite{morlet}. We first
have found the wavelengths for which the transparency $T(\lambda )$ of the
whole sample exceeds 0.5. For these wavelengths the spatial distributions of
the wave amplitude and of the amplitude of the corresponding wavelet
transformation have been compared. The examples are shown in Fig. 9. One can
see that in the areas where field is localized, i. e. in those that should
be transparent for the given frequency, the wavelet amplitude is strongly
suppressed, while in the non-transparent parts it has well pronounced
maxima. Therefore, in accordance with the resonant scattering mechanism, the
resonant cavity for a frequency $\omega $ arises in that area of a
disordered sample, where for some reason or other, the harmonic with the
wave number $q_{res}=2k$ in the power spectrum of the (random) refractive
index has small amplitude. In a sense the wavelet amplitude can be
interpreted as an effective potential. Note, not all resonance realizations
exhibit so good correlation with the wavelet analysis as shown in Fig. 9.
Actually the correlation was observed for 70 - 80 \% of all resonances.

\begin{figure}[htb]
\centering \includegraphics[width=80mm,height=100mm,angle=0,]{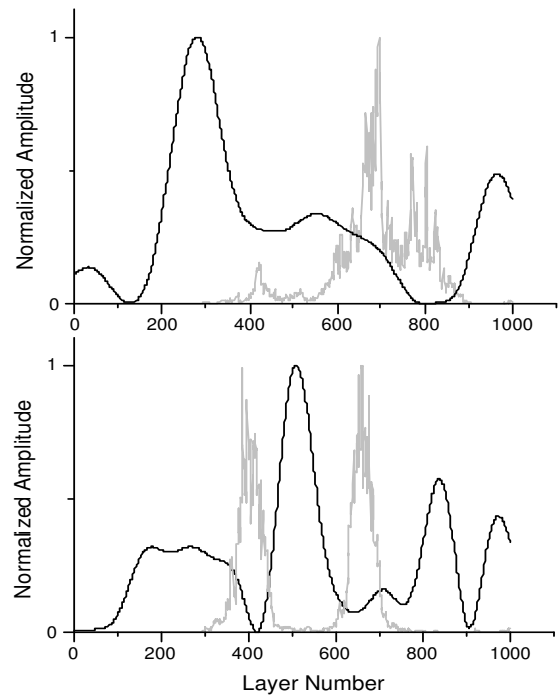}
\caption{Field amplitudes (light lines) and wavelet transformations (dark
lines) as functions of the coordinate in two random realizations.}
\end{figure}

\section{Conclusions}

From the results above it follows that the resonant transmission through a
disordered 1D system occurs due to the existence inside the system of a
transparent (for a given resonant frequency) segment with the size of order
of the localization length. The mode structure of the sample, the
transmission coefficient, and the intensity distribution at resonant
frequencies depend on the positions of the segment. These dependencies are
very robust, insensitive to the fine structure of the system, and therefore
can be described by corresponding formulas for an effective regular
potential profile. The fitting parameters are the total length of the sample
and the localization length, which is a self-averaging quantity and can be
found, for example, from the transmission coefficient at typical
(non-resonant) realizations. This feature enables to estimate the position
of an effective resonant cavity by measuring the transmission coefficient at
a typical and at the corresponding resonant frequencies. From the first data
the localization length can be obtained, the second one, $T_{res}$, could be
used to find the asymmetry parameter $d$ (i.e. the coordinate of the
effective cavity) from Eq. (16).Then Eq. (17) gives estimation for the
pumped intensity. Therefore, although \textit{locally} for 1D photons there
is no analog to the quantum mechanical tunnelling (by definition the
effective energy of 1D photons is always higher than the effective potential
barrier), \textit{macroscopically} (on scales larger than the localization
length) the problem of the propagation of light through a disordered 1D
system can be reformulated in terms of the effective potential profile.


\begin{thebibliography}{99}
\bibitem{lif} I. M. Lifshits, S. A. Gredeskul, and L. A. Pastur, \textit{%
Introduction to the Theory of Disordered Systems} (Wiley, New York, 1988).

\bibitem{berry} M. V. Berry, S. Klein, "Transparent mirrors: rays, waves and
localization", Eur. J. Phys. \textbf{18}, 222-228 (1997).

\bibitem{frish} U. Frisch, C. Froeschle, J.-P. Scheidecker, and P.-L. Sulem,
''Stochastic resonance in one-dimensional random media'', Phys. Rev. A 
\textbf{8}, 1416-1421 (1973).

\bibitem{azbel5} M. Ya. Azbel, P. Soven, ''Transmission resonances and the
localization length in one-dimensional disordered systems'', Phys. Rev. B 
\textbf{27}, 831-836, (1983).

\bibitem{azbel4} M. Ya. Azbel, ''Eigenstates and properties of random
systems in one dimension at zero temperature'', Phys. Rev. B \textbf{28},
4106-4125 (1983).

\bibitem{natu} D. S. Wiersma, ''The smallest random laser '', Nature \textbf{%
406}, 132-133 (2000).

\bibitem{natu2} D. S. Wiersma, S. Cavalieri, ''Light emission: A
temperature-tunable random laser '', Nature \textbf{414}, 708-709 (2001).

\bibitem{souc} Xunyia Jiang, C. M. Soucoulis, ''Time Dependent Theory for
Random Lazer'', Phys. Rev. Lett. \textbf{85, }70-73 (2000).

\bibitem{cao} H. Cao, Y. G. Zhao, S. T. Ho, E. W. Seelig, Q. H. Wang, and
R.P. H. Chang, ''Random Laser Action in Semiconductor Powder'', Phys. Rev.
Lett. \textbf{82}, 2278-2281 (1999).

\bibitem{cao2} H. Cao, Y. Ling, J. Y. Xu, and C. Q. Cao, and P. Kumar,
''Photon Statistics of Random Lasers with Resonant Feedback'', Phys. Rev.
Lett. \textbf{86}, 4524-4527 (2001).

\bibitem{bohm} D. Bohm, \textit{Quantum Theory} (Prentce-Hall, Inc, New
York, 1952).

\bibitem{ryt} S. Rytov, Yu. Kravtsov, and V. Tatarskii, \textit{Principles
of Statistical Radiophysics v. 4. Wave Propagation Through Random Media}
(Springer-Verlag, 1989).

\bibitem{morlet} A. Grossman, J. Morlet, "Decomposition of Hardy function
into square integrable wavelets of constant shape", SIAM J. Math. Anal. 
\textbf{15}, 273 (1984).
\end{thebibliography}
\end{document}